
\input phyzzx.tex
\date{January 12, 1994}
\voffset=3pc
\hoffset=0.45in

\def\PL  #1 #2 #3 {{\sl Phys.~Lett.}~{\bf#1} (#3) #2 }
\def\NP  #1 #2 #3 {{\sl Nucl.~Phys.}~{\bf#1} (#3) #2 }
\def\PR  #1 #2 #3 {{\sl Phys.~Rev.}~{\bf#1} (#3) #2 }
\def\PRD #1 #2 #3 {{\sl Phys.~Rev.~D} {\bf#1} (#3) #2 }
\def\PRB #1 #2 #3 {{\sl Phys.~Rev.~B} {\bf#1} (#3) #2 }
\def\PP  #1 #2 #3 {{\sl Phys.~Rep.}~{\bf#1} (#3) #2 }
\def\MPL #1 #2 #3 {{\sl Mod.~Phys.~Lett.}~{\bf#1} (#3) #2 }
\def\CMP #1 #2 #3 {{\sl Comm.~Math.~Phys.}~{\bf#1} (#3) #2 }
\def\PRL #1 #2 #3 {{\sl Phys.~Rev.~Lett.}~{\bf#1} (#3) #2 }
\def\TMP  #1 #2 #3 {{\sl Theor.~Math.~Phys.}~{\bf#1} (#3) #2 }
\def\JMP  #1 #2 #3 {{\sl Jour.~Math.~Phys.}~{\bf#1} (#3) #2 }
\def\IJ  #1 #2 #3 {{\sl Int.~Jou.~Mod.~Phys.}~{\bf#1} (#3) #2 }

\REF\wilson{K.~Wilson, \PRB 4 3185 1971 ; K. Wilson
and J. Kogut, {\sl Phys. Rep.}
{\bf 12} (1974) 75  .}
\REF\dyson{F. Dyson, \CMP 12 91 1969 ; G. Baker, \PRB 5 2622 1972 . }
\REF\sinai{P. ~Bleher and Y. ~Sinai, \CMP 45 247 1975 ; P.~Collet and
J. P. ~Eckmann,\CMP 55 67 1977 .}
\REF\gaw{K. Gawedzki
and A. Kupiainen, Les Houches 1985, K. Osterwalder and R. Stora, Editors }
\REF\preprints{Y. Meurice, Univ. of Iowa Preprint 92-12, hep-th 9307128. }
\REF\conje{Y. Meurice, \MPL 7 3331 1992 and  \JMP 35 0000 1994 .}
\REF\polyakov{A. ~Polyakov, \PL 82B 247  1979  and  \PL 103B 211 1981 . }

\REF\parisi{G. Parisi, {\it Statistical Field Theory} (Addison-Wesley,
New-York, 1988) and references therein. }
\REF\progress{Y. Meurice and G. Ordaz, work in progress.}
%

\Pubnum={U.Iowa 94-03 \cr
hep-lat/9401016 }
\titlepage
\title{ A Numerical Study of the Hierarchical Ising Model: High Temperature
Versus Epsilon Expansion}
\author{Y. Meurice, G. Ordaz, and V.G.J. Rodgers}
\address{Department of Physics and Astronomy, University of Iowa,
Iowa City, Iowa 52242, USA}
\vfil
\abstract
\def\powe{18}
We study numerically the magnetic susceptibility of the hierarchical model
with Ising
spins ($\sigma =\pm 1$) above the critical temperature and for two
values of the epsilon parameter.
The integrations are performed exactly using recursive methods which exploit
the symmetries of the model. Lattices with up to $2^{\powe }$ sites have been
used. Surprisingly, the numerical data can be
fitted very well with a simple power law of
the form $(1- \beta /\beta _c )^{- \gamma}$
for the {\it whole} temperature range.
The numerical values for $\gamma $ agree within a few percent with the values
calculated with a high-temperature expansion but show significant
discrepancies
with the epsilon-expansion.
\endpage

\chapter{Introduction}
The renormalization group (RG) method\refmark\wilson is a powerful
tool to handle critical phenomena and to approach the continuum limit
of lattice models. However, its practical implementation usually
requires
approximations. In his original paper, Wilson made order of
magnitude estimates of
various terms contributing to the partition
function of the Landau-Ginzburg model
and derived  the so called
approximate recursion formula.\refmark\wilson In this approximation,
the RG transformation
is reduced to a single integral equation which can be studied
 using numerical methods or functional analysis.

Recursion formulas closely related to the approximate recursion
formula hold {\it exactly}
for the hierarchical models.\refmark\dyson
This is due to the large group of symmetries of the hamiltonians of
these models.
The RG transformation for these models
has been studied in great detail and rigorous
results concerning the epsilon expansion of the critical exponents are
available
in the literature.\refmark\sinai

The fact that the RG transformation can be handled easily for hierarchical
models
suggests the use of these models as an approximation\refmark\gaw for nearest
neighbor models.
The main technical problems in proceeding this way are: how
to derive explicitly
the approximate models and how to improve systematically the approximation.
Recently, one of us\refmark\preprints has answered these questions for the
gaussian models where
everything can be calculated explicitly. In order to extend this method
to interacting models, one should be able to calculate the average value of
perturbation terms added to the hierarchical hamiltonian.
For Ising models, where the
spin $\sigma $ takes only the values $\pm 1$, this
task can be carried numerically
in an efficient way.
It has been suggested\refmark\conje that such an approach might shed some light
on Polyakov's conjecture\refmark\polyakov for the 3D Ising model.
In preparation for these calculations, we first checked the
agreement between the numerical calculations for the (unperturbed) hierarchical
Ising model and analytical results. In doing so, we found surprising
results which are reported in the following.

In this paper, we calculate numerically the magnetic susceptibility per site -
the susceptibility for short -
of the hierarchical
Ising model as a function of the temperature and for two values (0 and 1)
of $\epsilon $, the parameter used
in the $\epsilon$-expansion. Calculations have been carried
with up to $2^{\powe }$ sites.
The numerical integration made use of the symmetries of
the model in order to cut down the
time of computation logarithmically. However, no
approximations have been made and the
numbers shown below are exact up to round-off errors.
These errors were analyzed
by changing from simple to double precision. In all the cases considered, this
only affected the fifth significant digit
of the susceptibility in the worse cases.
Our calculations have been mostly restricted to the high temperature region
and its boundary. In other words, the parameter $\beta $,
proportional to the inverse
temperature, will run between 0 and a critical value $\beta _c$.
However, at the beginning,
a few calculations will be made in the low temperature region in order
to locate $\beta _c$.

Surprisingly, we found that the numerical data can be fitted very precisely
with a simple power law of
the form $(1- \beta /\beta _c )^{- \gamma}$
in the {\it whole} high-temperature region.
As a consequence, it is possible to calculate with good accuracy
the free entries of this
parametrization, $\gamma$ and $\beta _c$, using the two
first coefficients of the high-temperature
expansion of the susceptibility.
On the other hand, the values of $\gamma $ obtained
numerically, differ significantly from
the values obtained in the $\epsilon$-expansion .
This does not mean that the approximations
made to calculate the $\epsilon$-expansion are incorrect, but only
that for the models considered here,
we can never get near the region (in the space of theories) where these
approximations can be made.

This paper is organized as follows. In section 2, we introduce
the hierarchical model and the numerical method of integration.
In section 3, we check the scaling laws for the variance
of the total spin and we determine $\beta _c$.
In section 4, we analyze the temperature dependence of the susceptibility
and we show that the data can be fitted very well with a simple power law.
In section 5, we
compare the results with the high temperature expansion and the
$\epsilon$-expansion. Finally, we discuss our present understanding
of the results in the conclusions.

\chapter{The Hierarchical Ising Model}

In this section we describe the hierarchical Ising model and the basic
ideas of the numerical calculation
performed.
Hierarchical models\refmark\dyson
are specified by a non-local hamiltonian bilinear
in the spin variables and a local measure of integration.
We consider here the case of a Ising measure, where the spin
take only the values $\pm 1$.
The hamiltonian of a hierarchical model with $2^n$ sites can
be written as
$$H=-{1 \over 2} \sum\limits_{l=1}^{n}({c \over 4})^l
\sum\limits_{i_n,...,i_{l+1}}\ (\ \sum\limits_{i_l,....,i_1} \sigma
_{(i_n,....,i_1)} )^2\ \eqno(2.1) $$
For convenience, we have labeled the sites with $n$
indices $i_n ..... i_1$, each index
being 1 or 2. In order to visualize the meaning of
this notation, one can divide
the $2^n$ sites into two boxes, each containing $2^{n-1}$
sites. If $i_n =1$, the site is
the first box, if $i_n=2$, the site is in the second box.
Repeating this procedure $n$ times
(for the two boxes, their respective two sub-boxes, etc.),
we obtain an unambiguous labeling for each of the sites.
The interactions corresponding to a given value of $l$ in (2.1)
couple all the sites within each
box of size $2^l$ with the same strength $({c \over 4})^l $.
The model has a free parameter $c$ for which we shall use the
parametrization
$$c=2^{1-{2\over D}}.\eqno(2.2)$$
The parameter of the epsilon-expansion will be defined as
$$\epsilon = 4-D \eqno(2.3).$$
This choice has been justified in Ref.[\conje],
but different conventions exist in the literature.
In the actual calculations reported below, we have selected the
values 0 and 1 for $\epsilon $.

We are interested in calculating the magnetic susceptibility of the
hierarchical Ising model.
This quantity can be calculated easily if we know the probability for
the total spin denoted
$P_n(S)$. This probability is obviously $\beta $-dependent even though
we shall not write it explicitly.
This probability will be calculated recursively using the RG method
without a rescaling of the spins.
We first integrate the spins inside boxes of size 2 keeping the sum of
the spins in each box constant.
We then include the terms with $l=1$ in (2.1) in a new local measure for
these sums. We repeat
this procedure $n$ times and obtain a measure for the total spin which
can be normalized as probability.
Note that a choice of indices $i_n,......,i_{l+1}$ completely specifies
a box of size $2^l$ with
the subdivision described above. We call the sum of the spin inside this
box $S_{i_n,,....,i_{l+1}}$.
It can take all the even values between $-2^l$ and $2^l$.
Obviously, $S_{i_n,....,i_{l+1}}=S_{i_n,....,i_{l+1},1}+
S_{i_n,....,i_{l+1},2}$.
With these notations, the recursion formula reads
$$\eqalign{P_{l+1}(S_{i_n,.....,i_{l+1}})\ = \
&C_{l+1}\  Exp({1 \over 2}\beta
({c\over 4 })^{l+1}(S_{i_n,....,i_{l+1}})^2)\cr
. \sum_{ S_{i_n,....,i_{l+1}}=S_{i_n,....,i_{l+1},1}+S_{i_n,....,i_{l+1},2}}
&P_l(S_{i_n,....,i_{l+1},1})P_l(S_{i_n,....,i_{l+1},2})}  \eqno(2.4)$$

The constant $C_{l+1}$ is adjusted in such a way that the sum of
the probabilities add up to 1. Strictly speaking, it is not necessary
to impose such a normalization during the intermediate steps
of the calculation,
however it keeps the numbers reasonably small.
This recursion formula has been implemented with a computer program.
In order to calculate $P_n(S)$, we have to repeat $2^n$ times a calculation
involving roughly $2^n$ operations. Consequently,
the time necessary to calculate $P_n(S)$ scales approximately
like $4^n$. With the fastest computer at our disposal, a DEC alpha 3000/400,
it takes about 10 minutes to
calculate $P_{17}(S)$ from $P_{16}(S)$ when programmed in FORTRAN.

Due to the size of the calculation, it is clearly necessary to check for
round-off errors. We have studied the size of these errors by repeating
the calculation with double-precision instead of
simple precision for a large sample of values of $\beta $.
For $n$ up to 16, we have found very good agreement between the two
calculations,
the differences showing up in the sixth significant digit of the
susceptibility. Differences in the fifth digits were observed
for $n=17$ and 18.
Calculations for $n=19$ and beyond have shown less stability, require
a lot of computer time and will not be reported here.

The main quantity of interest for us is the average of the square of the total
spin denoted $X_n(\beta)$ and defined as
$$X_n(\beta)=\sum_S P_n(S) S^2 \eqno(2.5)$$
Above the critical temperature, this quantity divided by the number of sites
has a finite thermodynamic limit. It
will be called the magnetic
susceptibility per site and denoted $\chi _n (\beta) $
$$\chi _n (\beta)= {X_n (\beta) \over {2^n}} \eqno(2.6)$$

\chapter{The Scaling Laws and the Determination of $\beta _c$}

The RG method has definite predictions for the the large $n$ behavior
of $X_n$. This is explained at length for instance in chapter 7 of
Parisi's textbook.\refmark\parisi
In the following we shall just recall the main
results and show that our numerical calculations reproduce these results
with good precision.
The case $D=3$ will be discussed in complete detail, while the results for
$D=4$ will only be briefly commented upon.

It is convenient to express $X_n$ as a power of the square
of the number of sites, namely
$$X_n(\beta)\ = \ 2^{2n\omega(\beta , n)} \eqno(3.1)$$
The value of
\def\ome{\omega(\beta, n)}
$\ome$ can be easily estimated for large or small values of $\beta$.
For large values of $\beta$, all the spins tend to align and $\ome $
gets close to 1. For small values of $\beta$, the spins at different
sites become uncorrelated and the variance of the total spin
can be approximated by the sum of the individual variances. In other words
$\ome $ get close to $1/2$. A more detailed analysis\refmark\parisi shows that
for large $n$, $\ome$ is attracted by 1 if $\beta $
exceeds a critical value denoted $\beta_c$, and by $1/2$ if $\beta$ is less
than $\beta _c$. When $\beta$ is exactly $\beta _c$, $\ome $
tends to ${1 \over 2}
+ {1 \over D}$. This can be derived from the fact that
if we reabsorb a factor $\sqrt{c/4}=
2^{{1 \over 2}+ {1 \over D}}$
in the spin variable after the integration described in the previous section,
the Hamiltonian is invariant in the thermodynamic limit.

We shall now illustrate these results and calculate $\beta _c$ for $D=3$.
In order to get a rough idea of the value of $\beta _c$, we have plotted
in Fig. 1
the trajectories $\ome $ from $n=1$ to $n=16$ and
for $\beta = 0.1, 0.2,...,1.6$. It appears
clearly that the separation between the two
domains of attraction occurs for a value of $\beta $ between 1.1 and 1.2.
We have
repeated this calculation for $\beta =1.10, 1.11, ......,1.20$ as shown
on Fig. 2. This restricts $\beta _c$ to the interval $[1.17, 1.19]$.
At this point, it is more informative to consider ratios of susceptibilities
at two successive values of $n$, because this quantity depends less
sensitively on the constant of proportionality appearing in the
scaling laws. A short calculation shows that
$${\rm lim}_{n \rightarrow \infty }
Log_2 ({\chi _{n+1}(\beta _c )\over \chi _{n}(\beta _c )})\ =\
{2\over D} \eqno(3.3)$$
Fig. 3 displays $Log_2 ({\chi _{n+1}(\beta )\over \chi _{n}(\beta )})$ for
3 values of $\beta$ and for $n$ up to 18. For $\beta = 1.179$, this quantity
stays within one percent of the critical value $2/3$
for the last six iterations.
If $\beta$ is increased or decreased by $0.001$, the same quantity,
departs by about 10 percent
from $2/3$ when $n=18$. From this, we conclude that 1.179
approximates $\beta _c$ with a precision better than 0.001.

Applying the same procedure for $D=4$, we found that
$\beta _c$ should be inside the interval $[0.66,0.67]$.
Fig. 4 shows that 0.665 approximates
$\beta _c$ with a precision better than 0.001.

\chapter{Fitting the numerical data with a simple power law}

In this section, we report our numerical results
concerning the $\beta$-dependence of the magnetic
susceptibility per site at $D=3$ and $D=4$ and
for $\beta< \beta_c$.
Again, we start with the case $D=3$ and discuss it in detail while the case
$D=4$ will be presented  more rapidly  later in this section.

We have calculated $\chi_n(\beta)$ for $D=3$ and  $n$ up to 16 and
for values of $\beta$ between 0
and 1.18 separated by
intervals of length 0.01. As expected, the susceptibility rises sharply
when $\beta $ gets close
to 1.18. We have displayed the results for $n=16$ and $\beta \leq 1.1$
on Fig. 5. Results for
$n=14$ or $n=15$ would have been hardly distinguishable from $n=16$ on a
graph of this size.
On the other hand, the $n$-dependence becomes more sizable when $\beta $
is closer to its critical value
as shown on Fig. 6 which also includes  results for $n=17$.
Except for the difference of scales, the resemblance between
Fig. 5 and Fig. 6 is striking.
This can be understood from the fact that
the numerical data can be fitted very precisely
with a simple power law of
the form
$$\chi_n(\beta)\ =\ (1- \beta /\beta _c )^{- \gamma}\eqno(4.1)$$
in the {\it whole} interval $[0, 1.18]$.
In order to justify this claim, we first notice that (4.1)
implies the inverse logarithmic derivative of the susceptibility
is a linear function, namely
$$({d\over{d\beta}} Log(\chi_n(\beta)))^{-1}\ =\ (\gamma )^{-1} (\beta _c
- \beta ) \eqno(4.2)$$
We now approximate this inverse logarithmic derivative by
$$\Delta \beta / \Delta Log ( \chi _n  (\beta +
{{\Delta \beta}/2}))\ = \  {{\Delta \beta} \over {
Log(\chi _n (\beta + \Delta \beta ))-Log(\chi _n (\beta ))}}\eqno(4.3)$$
with $\Delta \beta =0.01$, the interval used here.
This function is plotted in Fig. 7 for $n=16$. Remarkably, the numerical data
is barely distinguishable from the least square linear fit
$$\Delta \beta / \Delta Log ( \chi _{16}  (\beta ))\
= 0.80839-0.67843 \beta \eqno(4.4)$$

The graphs for n=14 or 15 are almost identical. The linear
fits are respectively
$$\eqalign{
\Delta \beta / \Delta Log ( \chi _{15}  (\beta))\
=& 0.80849-0.67712 \beta \cr
\Delta \beta / \Delta Log ( \chi _{14} (\beta ))\
=& 0.80867-0.67508 \beta }\eqno(4.5)$$

In order to see the corrections to the linear behavior, we have plotted
in Fig.8 the difference between the linear fit and the data,
denoted $E(\beta )$, for $n=\ $ 14, 15 and 16.
These difference are not larger than 0.003 in absolute values and have
interesting
regularities.
The reasons why the power law (4.1) is so accurate and the
nature of its corrections are being investigated.

We have intentionally used simple precision data to plot Fig. 8 in order
to give an idea of the round-off errors of the method. Small irregularities
are visible especially in the low $\beta $ region where their typical size
is $10^{-4}$. On the other hand, $E(\beta )$ is smoother for $\beta > 0.8$.
The size of these numerical errors is compatible with
the claim made in section
that numerical errors affects only the sixth significant digit of the
susceptibility. Indeed, if one of the $\chi_n $
is replaced by $\chi_n (1+ \delta )$
in Eq. (4.3), this creates an error
$\delta \Delta \beta /(\Delta Log(\chi))^2$. A simple inspection shows
that $\delta $ can be amplified by a factor of order 100 if $\beta $ is not
too close to $\beta _c$.

A similar procedure has been followed for $D=4$.
In Fig. 9 we desplay the quantity
$\Delta \beta / \Delta Log ( \chi _{16}  (\beta ))\ $.
Again the departures from linearity are small and the
changes in the coefficients
of the linear fit for $n=14$ and 15 are of the same order as those for
$D=3$. The difference between the linear fit and the data for $n=$ 14, 15
and 16 is displayed on Fig.10.

\chapter{Comparisons with High-Temperature and Epsilon Expansions}

Approximate values of the critical
exponent $\gamma $ can be extracted from linear fits presented
in the last section. For instance,
identifying Eqs.(4.2) and (4.4) yields $\gamma = 1.474$ and $\beta _c = 1.192$.
Changes of 1 or 2 percent can be obtained if we restrict $\beta $ to
intervals closer to $\beta _c$.
These results can be compared with analytical results obtained using the high
temperature or the $\epsilon$-expansion.

The high-temperature expansion of the susceptibility reads
$$\chi_n (\beta)=1\ + \ b_{(1,n)}\beta \ + \ b_{(2,n)}
\beta ^2 \ + \ ... \eqno(5.1)$$
A straightforward but tedious calculation yields
\def\cf{{c\over 4}}
\def\ct{{c\over 2}}
\def\ce{{{c^2}\over 8}}
$$b_{(1,n)}=(1-\cf )^{-1} (\cf (1-(\ct)^n )
(1-\ct)^{-1} - (2^n -1)(\cf )^{n+1})\eqno(5.2.1)$$
and
$$\eqalign{b_{(2,n)}=&(b_{(1,n)})^2 - (1-\cf )^{-2}
[(\cf)^2 (1-(\ce)^n ) (1-\ce)^{-1} \cr
&-2 (\cf )^{n+2}(1-(\ct)^n ) (1-\ct)^{-1}
- (2^n -1)(\cf )^{2(n+1)}]}
\eqno(5.2.2)$$
Comparing Eq.(5.1) and Eq.(4.2) expanded about 0, we obtain
$$\gamma \ = \ ({{2b_{(2,n)}}\over{b_{(1,n)}^2}}-1)^{-1}\eqno(5.3)$$

On the other hand, the critical exponents of the hierarchical model
near Wilson's non-trivial fixed point can be calculated using the $\epsilon$-
expansion. With the convention of Eq.(2.2), we obtain
at first order in $\epsilon $.
$$\gamma = 1+0.1667 \epsilon \eqno(5.4)$$

These analytical results are compared with the numerical results in Fig.11
and 12 for $D=3$ and $D=4$ respectively. It appears clearly that the data
agree within a few percent with
the high-temperature calculation performed above, while a
significant departure from the
$\epsilon$-expansion is shown. This is quite clear in the case $D=4$ where
the $\epsilon $ expansion predicts $\gamma =1$. This is less clear in the
case $D=3$ where we do not know at which order we need to truncate what
is presumably an asymptotic series.

\chapter{Conclusions}

We have studied the magnetic susceptibility per site of the hierarchical
Ising model. We found that the $\beta $-dependence of this quantity
could be fitted quite accurately with a simple power law.
It would be quite interesting to understand the origin of this result.
Small corrections to this power law have been observed and it might
be possible to handle this problem using Callan-Symanzik's equations.
In any case, it is remarkable that one is able to obtain an
accurate information about the critical behavior of a model just by
calculating two coefficients of a high temperature expansion.

We insist on the fact that we have made no approximations in our numerical
calculations and that the results presented are exact up to round-off
errors. We have analyzed these errors and found that they do not affect our
conclusions. The fact that we were able to reproduce
accurately well-understood
analytical results such as the scaling laws at the critical temperature and
the high temperature behavior of the susceptibility seem to rule out
errors in implementing Eq. (2.4) numerically.

The discrepancy with the $\epsilon$-expansion is quite surprising.
Our limited data cannot rule out the possibility that this discrepancy
decreases very slowly with $n$. Another possibility is that by starting  with
Ising spins and applying the RG transformation, we never get close enough
to the region (in the space of theories) where the $\epsilon$-expansion is
legitimate. This second possibility seems supported by preliminary
results\refmark\progress obtained in the field theory approach of this problem

\ack
One of us (Y.M.) would to thank the participants of the
Aspen Center for Physics in June and July 1993 for providing
a stimulating atmosphere while some initial motivations for this
work where found.
This work was partially supported by the Carver foundation and
NSF grant PHY-9103914.
\refout
\vfill
\eject
\centerline{FIGURE CAPTIONS}

\noindent
Fig. 1 : $\omega (n ,\beta)$ versus $n$ for $D=3$ and values of $\beta $
going from
0.1 to 1.6 by steps of 0.1

\noindent
Fig. 2 :  $\omega (n ,\beta)$ versus $n$ for $D=3$ and values of $\beta $
going from
1.1 to 1.2 by steps of 0.01

\noindent
Fig. 3 : $Log_2$ of the ratio of two successive values of the susceptibility
for $D=3$ and for $\beta $= 1.178, 1.179 and 1.180.

\noindent
Fig. 4 : $Log_2$ of the ratio of two successive values of the susceptibility
for $D=4$ and for $\beta $= 0.6645, 0.6655 and 0.6665 .

\noindent
Fig. 5 : The magnetic susceptibility for $D=3$, $n=16$ and values of $\beta $
going from
0 to 1.18 by steps of 0.01.

\noindent
Fig. 6 : The magnetic susceptibility for $D=3$, $n=$14, 15, 16 and 17,
and values of $\beta $
going from
1.00 to 1.18 by steps of 0.01.

\noindent
Fig. 7 : A discrete version of the inverse logarithmic derivative of the
susceptibility and its linear fit for $D=3$ and $n=16$.

\noindent
Fig. 8 : Difference between the discrete version of the inverse
logarithmic derivative of the
susceptibility and its linear fit for $D=3$ and $n=14$, 15 and 16.

\noindent
Fig. 9 : A discrete version of the inverse logarithmic derivative of the
susceptibility and its linear fit for $D=4$ and $n=16$.

\noindent
Fig. 10  : Difference between the discrete version of the inverse
logarithmic derivative of the
susceptibility and its linear fit for $D=4$ and $n=14$, 15 and 16.

\noindent
Fig. 11 : Comparison between the high temperature calculation and our
numerical estimation
of the critical exponent $\gamma $ for $D=3$ and $n$ between 5 and 16.
The solid line
is the result for the $\epsilon $-expansion at first order.

\noindent
Fig. 12 : Comparison between the high temperature calculation and
our numerical estimation
of the critical exponent $\gamma $ for $D=4$ and $n$ between 5 and 16.
The solid line
is the result for the $\epsilon $-expansion.
\end